\documentclass[aps,prl,superscriptaddress,showpacs,showkeys,floatfix,twocolumn]{revtex4}

  \usepackage[dvips]{graphicx}

\def\NIM{Nucl. Instrum. Methods~}
\def\NIMA{{Nucl. Instrum. Methods}~{\bf A}}

\def\NPB{{Nucl. Phys.}~{\bf B}}
\def\PLB{{Phys. Lett.}~{\bf B}}

\def\PRL{Phys. Rev. Lett.\ }
\def\PRD{{Phys. Rev.}~{\bf D}}

\begin{document}

\title{Double Helicity Asymmetry in Inclusive Mid-Rapidity \\
$\pi^{0}$ Production for Polarized $p+p$ Collisions at $\sqrt{s}=200$~GeV}

\newcommand{\abilene}{Abilene Christian University, Abilene, TX 79699, USA}
\newcommand{\acadsin}{Institute of Physics, Academia Sinica, Taipei 11529, Taiwan}
\newcommand{\banaras}{Department of Physics, Banaras Hindu University, Varanasi 221005, India}
\newcommand{\barc}{Bhabha Atomic Research Centre, Bombay 400 085, India}
\newcommand{\bnl}{Brookhaven National Laboratory, Upton, NY 11973-5000, USA}
\newcommand{\caucr}{University of California - Riverside, Riverside, CA 92521, USA}
\newcommand{\ciae}{China Institute of Atomic Energy (CIAE), Beijing, People's Republic of China}
\newcommand{\cns}{Center for Nuclear Study, Graduate School of Science, University of Tokyo, 7-3-1 Hongo, Bunkyo, Tokyo 113-0033, Japan}
\newcommand{\colorado}{University of Colorado, Boulder, CO 80309}
\newcommand{\columbia}{Columbia University, New York, NY 10027 and Nevis Laboratories, Irvington, NY 10533, USA}
\newcommand{\dapnia}{Dapnia, CEA Saclay, F-91191, Gif-sur-Yvette, France}
\newcommand{\debrecen}{Debrecen University, H-4010 Debrecen, Egyetem t{\'e}r 1, Hungary}
\newcommand{\elte}{ELTE, E{\"o}tv{\"o}s Lor{\'a}nd University, H - 1117 Budapest, P{\'a}zm{\'a}ny P. s. 1/A, Hungary}
\newcommand{\fsu}{Florida State University, Tallahassee, FL 32306, USA}
\newcommand{\gsu}{Georgia State University, Atlanta, GA 30303, USA}
\newcommand{\hiroshima}{Hiroshima University, Kagamiyama, Higashi-Hiroshima 739-8526, Japan}
\newcommand{\ihepprot}{Institute for High Energy Physics (IHEP), Protvino, Russia}
\newcommand{\illuiuc}{University of Illinois at Urbana-Champaign, Urbana, IL 61801}
\newcommand{\isu}{Iowa State University, Ames, IA 50011, USA}
\newcommand{\jinrdubna}{Joint Institute for Nuclear Research, 141980 Dubna, Moscow Region, Russia}
\newcommand{\kek}{KEK, High Energy Accelerator Research Organization, Tsukuba-shi, Ibaraki-ken 305-0801, Japan}
\newcommand{\kfki}{KFKI Research Institute for Particle and Nuclear Physics (RMKI), H-1525 Budapest 114, POBox 49, Hungary}
\newcommand{\korea}{Korea University, Seoul, 136-701, Korea}
\newcommand{\kurchatov}{Russian Research Center ``Kurchatov Institute", Moscow, Russia}
\newcommand{\kyoto}{Kyoto University, Kyoto 606-8394, Japan}
\newcommand{\labllr}{Laboratoire Leprince-Ringuet, Ecole Polytechnique, CNRS-IN2P3, Route de Saclay, F-91128, Palaiseau, France}
\newcommand{\lawllnl}{Lawrence Livermore National Laboratory, Livermore, CA 94550, USA}
\newcommand{\losalamos}{Los Alamos National Laboratory, Los Alamos, NM 87545, USA}
\newcommand{\lpc}{LPC, Universit{\'e} Blaise Pascal, CNRS-IN2P3, Clermont-Fd, 63177 Aubiere Cedex, France}
\newcommand{\lund}{Department of Physics, Lund University, Box 118, SE-221 00 Lund, Sweden}
\newcommand{\muenster}{Institut fuer Kernphysik, University of Muenster, D-48149 Muenster, Germany}
\newcommand{\myongji}{Myongji University, Yongin, Kyonggido 449-728, Korea}
\newcommand{\nagasaki}{Nagasaki Institute of Applied Science, Nagasaki-shi, Nagasaki 851-0193, Japan}
\newcommand{\newmex}{University of New Mexico, Albuquerque, NM 87131, USA}
\newcommand{\nmsu}{New Mexico State University, Las Cruces, NM 88003, USA}
\newcommand{\ornl}{Oak Ridge National Laboratory, Oak Ridge, TN 37831, USA}
\newcommand{\orsay}{IPN-Orsay, Universite Paris Sud, CNRS-IN2P3, BP1, F-91406, Orsay, France}
\newcommand{\peking}{Peking University, Beijing, People's Republic of China}
\newcommand{\pnpi}{PNPI, Petersburg Nuclear Physics Institute, Gatchina, Russia}
\newcommand{\riken}{RIKEN (The Institute of Physical and Chemical Research), Wako, Saitama 351-0198, JAPAN}
\newcommand{\rkrbrc}{RIKEN BNL Research Center, Brookhaven National Laboratory, Upton, NY 11973-5000, USA}
\newcommand{\saopaulo}{Universidade de S{\~a}o Paulo, Instituto de F\'{\i}sica, Caixa Postal 66318, S{\~a}o Paulo CEP05315-970, Brazil}
\newcommand{\seoulnat}{System Electronics Laboratory, Seoul National University, Seoul, South Korea}
\newcommand{\stonybrkc}{Chemistry Department, Stony Brook University, Stony Brook, SUNY, NY 11794-3400, USA}
\newcommand{\stonycrkp}{Department of Physics and Astronomy, Stony Brook University, SUNY, Stony Brook, NY 11794, USA}
\newcommand{\subatech}{SUBATECH (Ecole des Mines de Nantes, CNRS-IN2P3, Universit{\'e} de Nantes) BP 20722 - 44307, Nantes, France}
\newcommand{\tenn}{University of Tennessee, Knoxville, TN 37996, USA}
\newcommand{\titech}{Department of Physics, Tokyo Institute of Technology, Tokyo, 152-8551, Japan}
\newcommand{\tsukuba}{Institute of Physics, University of Tsukuba, Tsukuba, Ibaraki 305, Japan}
\newcommand{\vandy}{Vanderbilt University, Nashville, TN 37235, USA}
\newcommand{\waseda}{Waseda University, Advanced Research Institute for Science and Engineering, 17 Kikui-cho, Shinjuku-ku, Tokyo 162-0044, Japan}
\newcommand{\weizmann}{Weizmann Institute, Rehovot 76100, Israel}
\newcommand{\yonsei}{Yonsei University, IPAP, Seoul 120-749, Korea}
\affiliation{\abilene}
\affiliation{\acadsin}
\affiliation{\banaras}
\affiliation{\barc}
\affiliation{\bnl}
\affiliation{\caucr}
\affiliation{\ciae}
\affiliation{\cns}
\affiliation{\colorado}
\affiliation{\columbia}
\affiliation{\dapnia}
\affiliation{\debrecen}
\affiliation{\elte}
\affiliation{\fsu}
\affiliation{\gsu}
\affiliation{\hiroshima}
\affiliation{\ihepprot}
\affiliation{\illuiuc}
\affiliation{\isu}
\affiliation{\jinrdubna}
\affiliation{\kek}
\affiliation{\kfki}
\affiliation{\korea}
\affiliation{\kurchatov}
\affiliation{\kyoto}
\affiliation{\labllr}
\affiliation{\lawllnl}
\affiliation{\losalamos}
\affiliation{\lpc}
\affiliation{\lund}
\affiliation{\muenster}
\affiliation{\myongji}
\affiliation{\nagasaki}
\affiliation{\newmex}
\affiliation{\nmsu}
\affiliation{\ornl}
\affiliation{\orsay}
\affiliation{\peking}
\affiliation{\pnpi}
\affiliation{\riken}
\affiliation{\rkrbrc}
\affiliation{\saopaulo}
\affiliation{\seoulnat}
\affiliation{\stonybrkc}
\affiliation{\stonycrkp}
\affiliation{\subatech}
\affiliation{\tenn}
\affiliation{\titech}
\affiliation{\tsukuba}
\affiliation{\vandy}
\affiliation{\waseda}
\affiliation{\weizmann}
\affiliation{\yonsei}
\author{S.S.~Adler}	\affiliation{\bnl}
\author{S.~Afanasiev}	\affiliation{\jinrdubna}
\author{C.~Aidala}	\affiliation{\columbia}
\author{N.N.~Ajitanand}	\affiliation{\stonybrkc}
\author{Y.~Akiba}	\affiliation{\kek} \affiliation{\riken}
\author{A.~Al-Jamel}	\affiliation{\nmsu}
\author{J.~Alexander}	\affiliation{\stonybrkc}
\author{K.~Aoki}	\affiliation{\kyoto}
\author{L.~Aphecetche}	\affiliation{\subatech}
\author{R.~Armendariz}	\affiliation{\nmsu}
% ...
\collaboration{PHENIX Collaboration} \noaffiliation

\date{\today}

\begin{abstract}
We present a measurement of the double longitudinal spin asymmetry in
inclusive $\pi^{0}$ production in polarized proton-proton collisions at
$\sqrt{s}=200$ GeV.  The data were taken at the Relativistic Heavy Ion
Collider with average beam polarizations of 27\%.  The measurements are
the first of a program to study the longitudinal spin structure of the
proton, using strongly interacting probes, at collider energies.  The
asymmetry is presented for transverse momenta 1-5 GeV/$c$ at mid-rapidity,
where next-to-leading order perturbative quantum chromodynamic (NLO pQCD)
calculations well describe the unpolarized cross section.  The observed
asymmetry is small and is compared with a NLO pQCD calculation with a
range of polarized gluon distributions.
\end{abstract}




\maketitle  

From polarized lepton-nucleon deep inelastic scattering (DIS) experiments
over the past 20 years it is known that only $\sim 25\%$ of the proton
spin is carried by quarks and anti-quarks~\cite{DISsigma}. The rest of the
proton spin must hence be carried by the gluons and orbital angular
momentum.  DIS experiments have constrained the possible gluon
polarization in the proton through the measurement of scaling violation in
inclusive polarized scattering~\cite{DISg1}, and through semi-inclusive
measurements of two hadrons to utilize the photon-gluon fusion
process~\cite{DIShad}.  A fixed target experiment at Fermilab first
presented a measurement with strongly interacting probes~\cite{E704}.  
The reach of these measurements was limited, due to the low energy
available for fixed target experiments.  Presently, the gluon contribution
to the proton spin is largely unknown.

The polarized proton collisions at the Relativistic Heavy Ion Collider 
(RHIC) provide a new laboratory to study the proton spin structure with
strongly interacting probes.  
The PHENIX experiment has reported the unpolarized cross section 
for $\pi^0$ production at mid-rapidity for $p_T$=1-14 GeV/$c$, 
which is described well by next-to-leading-order perturbative QCD (NLO pQCD) 
calculations over 8 orders of magnitude~\cite{pi0_prl}.
In this Letter we report the first results on the double
spin asymmetry, $A_{LL}$, for inclusive $\pi^0$ production at mid-rapidity 
in longitudinally polarized proton-proton collisions
corresponding to $0.22$~pb$^{-1}$ integrated luminosity
with the PHENIX detector.

In perturbative QCD $A_{LL}$ is directly sensitive to
the polarized gluon distribution function in the proton
through gluon-gluon and gluon-quark subprocesses~\cite{bunce}.

The double spin asymmetry in $\pi^{0}$ production is given by
\begin{equation}
A_{LL}^{\pi^{0}} = \frac{\sigma_{++} -
\sigma_{+-}}{\sigma_{++}+\sigma_{+-}}
\end{equation}
where $\sigma_{++} (\sigma_{+-})$ is the cross section of the reaction
when two colliding particles have the same (opposite) helicity. 
Here we neglect the parity violating difference in cross section between 
$(++) \leftrightarrow (--)$ and $(+ -)  \leftrightarrow (- +)$ beam helicity 
configurations. Since the cross section can be obtained by dividing the 
experimental yield ($N$) by the integrated luminosity ($L$), 
$A_{LL}$ is expressed as
\begin{equation}
A_{LL} = \frac{1}{|\langle P_{B}P_{Y} \rangle|} \cdot \frac{N_{++}-R\cdot
N_{+-}}{N_{++} + R \cdot N_{+-}};~~~R=\frac{L_{++}}{L_{+-}},
\label{eq:a_ll}
\end{equation}
where $P_{Y(B)}$ are the polarizations of the RHIC ``yellow'' (``blue'') 
beams, and $R$ is the ratio of luminosities of protons
colliding with like to unlike helicities.  

For the 2002-2003 RHIC run, 55 bunches of polarized protons, 
typically $5\times10^{10}$ protons per bunch, were loaded into each 
of the yellow and blue counter-rotating accelerator/storage 
rings of RHIC and accelerated to 100 GeV. The bunch lengths and 
separations were $\sim$1 ns and 213 ns, respectively.
The beam polarization sign for each bunch 
was prepared independently at the source, with the
successive bunches in one ring alternating in polarization sign, 
and with successive pairs of bunches in the other ring alternating in sign.
The locations of the 
bunches were identified relative to a RHIC timing clock.
In this way, the experiments
collected data from collisions with all four combinations of blue-yellow 
ring beam polarization signs simultaneously. 

The stable direction of the proton spin in 
RHIC is vertical, but the region around the PHENIX 
experiment includes sets of magnets (spin rotators) to rotate the spin 
to the longitudinal direction at 
the collision point, and then back to vertical after the interaction point, 
in order to provide collisions with longitudinal polarization, and to maintain 
the required vertical polarization around RHIC.  The RHIC polarimeters measure 
the transverse beam polarization away from the interaction points, independent 
of the operation of the spin rotators. 

The transverse beam polarization was measured in RHIC independently 
in each beam using
proton-carbon elastic scattering in the coulomb nuclear interference (CNI)
region~\cite{pol}.  
The analyzing power $A_{N}^{pC}$ was measured for 22 GeV beam energy, 
$A_{N}^{pC}(22)$, to $\pm$30\%~\cite{tojo}. 
The energy dependence of the analyzing power over the RHIC energies is 
expected to be small, $<$10\%~\cite{trueman}.
For the results reported here, we have used the same analyzing power 
at 100 GeV as at 22 GeV, and 10\% is added in quadrature
to $\delta A_{N}^{pC}(22)$ to give a $\pm 32\%$ uncertainty for 
$A_{N}^{pC}(100)$. With these assumptions, 
the average polarization 
in the analyzed data set in this paper was 
$\sqrt{\langle P_B P_Y \rangle}=[27.0 \pm 0.3 (stat) \pm 3(syst)
\pm 8(A_{N}^{pC}~syst)]\%$.

Local polarimeters, sensitive to the 
transverse polarization at collision, were used to set up the spin rotators, 
and to monitor the beam polarization direction at the PHENIX experiment.  
The local polarimeters utilized a transverse single spin asymmetry in 
neutron production in $p$-$p$ collisions 
at $\sqrt{s}=200$ GeV~\cite{locpol}.   For vertically polarized beam a 
left-right asymmetry is observed for neutrons 
produced at very forward angles, with no asymmetry for production 
at very backward angles.  A fully longitudinally polarized beam produces 
no asymmetry.


Neutrons with $E_{n} > 20 $ GeV and production angle $0.3 < \theta_{n} < 2.5$ mrad
were observed by two hadronic calorimeters located $\pm 18$ meters from the
interaction point (ZDC or Zero Degree Calorimeter~\cite{nim_zdc}) . Scintillator hodoscopes
at 1.7 interaction length provided the neutron position at the ZDC, and thus the neutron 
production angle and azimuthal angle $\phi = \arctan (x/y)$ with $\hat y$ vertically upward. 
The $\hat x$ axis forms a right handed coordinate system with the $\hat z$ axis defined by the beam
direction for forward production.  
The single-spin asymmetry $\epsilon$ was calculated versus azimuth, 
from the four rates $N_{\uparrow,\phi}$,
$N_{\uparrow,\phi +\pi}$,$N_{\downarrow,\phi}$,$N_{\downarrow,\phi +\pi}$, 
using the geometric mean~\cite{sqrtform}.  This method largely 
cancels differences in luminosity between $\uparrow$ and $\downarrow$ 
polarization collisions and between detector acceptance differences 
at $\phi$ and $\phi +\pi$. 
Fig.~1 shows the observed asymmetry, for the spin rotators off and on, 
for the blue and yellow beams.  
With the spin rotators off, a left-right 
asymmetry is observed from the vertically polarized beam.  
With the spin rotators on, the measured transverse polarization, averaged 
over the run, was 
$\langle P_{Bx} \rangle =  3.3\% \pm 1.9\%$, 
$\langle P_{By} \rangle =  0.8\% \pm 2.0\%$, 
$\langle P_{Yx} \rangle = -2.0\% \pm 1.3\%$ and 
$\langle P_{Yy} \rangle =  5.4\% \pm 1.7\%$, out of $\langle P \rangle=27\%$.
The double spin transverse polarization was 
$\langle P_{Bx} P_{Yx} \rangle = ( 0.4 \pm 1.1) \cdot 10^{-3}$ and 
$\langle P_{By} P_{Yy} \rangle = (-0.2 \pm 0.8) \cdot 10^{-3}$, compared to 
$\langle P_{B} P_{Y} \rangle = 0.07$.
Therefore, with the spin rotators on, the transverse 
asymmetry is greatly reduced, indicating a high degree of 
longitudinal polarization. 


A separate run with the spin rotators set to 
give radial polarization confirmed the direction of 
the polarization for each beam.

\begin{figure}
\includegraphics[width=1.0\linewidth]{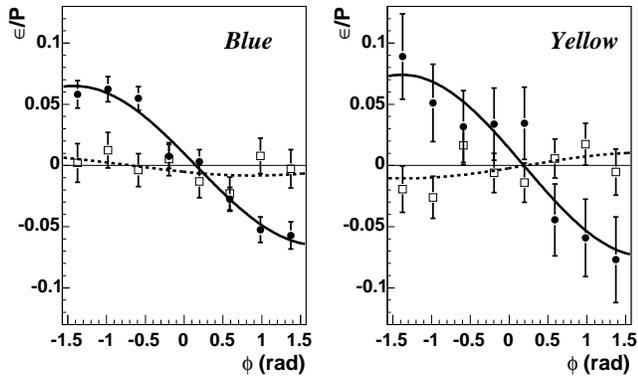}
\caption{\label{fig:pol}  Single spin raw asymmetry normalized by 
the beam polarization, $\epsilon/P$, as a 
function of azimuthal angle $\phi$, for forward neutron production. 
The solid points and the curve correspond to the spin rotators off 
(transverse polarization) condition and the open points and dashed
curve correspond to the spin rotators on (longitudinal polarization). 
Curves are sine function fits to the data, representing possible transverse
polarization. The data are for individual runs, where the blue (yellow) 
polarization was 24\% and 33\% (8\% and 28\%), for spin rotators off and on, 
correspondingly.}
\end{figure}

Collisions in PHENIX are defined by the coincidence of signals in two 
beam-beam counters (BBC)~\cite{nim_bbc} located at $\pm$1.44~m from the 
nominal interaction point and subtending a pseudo-rapidity range 
$\pm (3.0-3.9)$ with full azimuthal coverage. The BBCs select about
half of the 
inelastic proton-proton collisions~\cite{pi0_prl}.
The vertex was reconstructed from the time difference of the hits 
in the two BBCs. 
The collision vertex was required to be within 30 cm of the nominal 
interaction point. Events satisfying this condition constitute the minimum
bias (MB) trigger, which was used for relative luminosity measurements. 

A coincidence of the two ZDCs was used to estimate the possible bias in the 
relative luminosity measurement from the BBCs. 
This was done by comparing the accumulated number of triggers
in the ZDCs and BBCs for each bunch and each fill.
The accuracy of relative luminosity measurements $\delta R$  (Eq.~2)
was estimated to be $2.5 \times 10^{-4}$, which for the average beam 
polarization of 27\%, translated to $\delta A_{LL}=1.8 \times 10^{-3}$, 
and, on the same uncertainty level, confirmed no $A_{LL}$ asymmetry of 
BBC triggers relative to ZDC. The ratio $R$ averaged over the data 
sample used in the analysis was within 0.5\% of unity.

Neutral pions were reconstructed from the $\pi^{0}\rightarrow \gamma \gamma$
decays using finely granulated 
($\Delta \phi \! \times \! \Delta \eta \sim 0.01 \! \times \! 0.01$) 
electromagnetic calorimeters (EMCal)~\cite{nim_emc}, which consisted of 
two subsystems: a lead scintillator (PbSc) and a lead glass (PbGl) calorimeter
covering three quarters and one quarter of the EMCal acceptance, 
respectively.
Located at a radial distance of $\sim$5~m 
from the beam line, the EMCal covered the pseudorapidity 
range of $|\eta| < 0.35$ and two azimuthal angle intervals 
of $\Delta\phi \! \approx \! 90^{\circ}$ separated by 
$\phi \approx 70^{\circ}$ (nearly back-to-back).

High $p_T$ $\pi^0$'s were collected using coincidences between a MB 
trigger and an EMCal-based high $p_T$ photon trigger~\cite{pi0_prl}. 
The trigger efficiency for $\pi^0$'s varied from 8\% in the 1-2 GeV/$c$ 
$p_T$ bin to 90\% in the 4-5 GeV/$c$ $p_T$ bin. 

\begin{figure}
\includegraphics[width=0.9\linewidth]{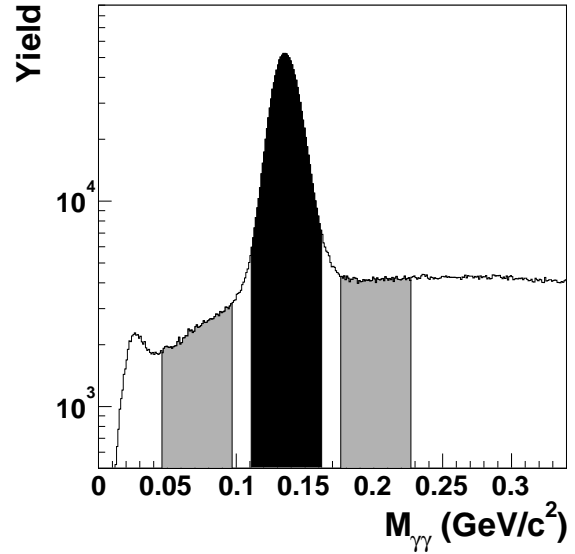}
\caption{\label{fig:mass} 
Two photon invariant mass, $M_{\gamma \gamma}$, distribution for 
the $p_T$ = 2-3 GeV/$c$ range.  The $\pi^0$ signal was selected 
for $\pm$25 MeV from the $\pi^0$ mass, indicated in black.  
The grey sidebands were used to estimate
the $A_{LL}$ of the background under the $\pi^0$ peak.
}
\end{figure}

The $\pi^0$ reconstruction and photon identification cuts were optimized 
to minimize the background contribution under the $\pi^0$ peak in the 
invariant mass distribution while keeping the $\pi^0$ efficiency high. 
For photon identification we used the shower profile and the time-of-flight 
measured by the EMCal,
and charge veto cuts. 
The charge veto was set for those EMCal 
clusters associated with hit(s) in the pad 
chamber~\cite{nim_pc}, which was 
located $\sim$ 20 cm in front of the EMCal surface.
In order to avoid the effects of electronic noise and suppress the 
very low energy background, only clusters with energy greater 
than 0.1 GeV in PbSc and 0.2 GeV in PbGl were used in the analysis. 

The $\pi^0$ yield was extracted by integrating the two photon 
invariant mass spectrum
over a $\pm$25~MeV/$c^2$ region around the $\pi^0$ mass
(signal region) as shown in Fig.~2 by the dark band.
The EMCal resolution was such that the widths of the $\pi^0$ 
mass peaks varied from 12~MeV/$c^2$ in 1-2~GeV/$c$ $p_T$ bin to 9.5~MeV/$c^2$ 
in 4-5~GeV/$c$ $p_T$ bin, in both PbSc and PbGl.
In the $p_T$ range of 1 to 5 GeV/$c$, 4 million $\pi^{0}$ candidates were
collected.
The background contribution (combinatorial+hadronic) under the 
$\pi^{0}$ peak, $r$, varied from 27\% 
in the 1-2 GeV/$c$ bin to 8\% in the 4-5 GeV/$c$ bin, as indicated in 
Table~\ref{tab:final}. 
The $\pi^{0}$ reconstruction efficiency due to photon identification cuts 
varied from 84\% in the lowest $p_T$
bin to $93\%$ in the highest $p_T$ bin.


The asymmetry of the background in the signal region, $A_{LL}^{BG}$, 
was evaluated
using the asymmetry calculated from the grey bands on both sides of the
signal (Fig.~2). 
The measured $\pi^0$ asymmetry, $A_{LL}^{raw}$, was corrected for 
the contribution 
of background using
\begin{equation}
A^{\pi^0}_{LL} = \frac{A_{LL}^{raw}-rA^{BG}_{LL}}{1-r},~
\sigma_{A^{\pi0}_{LL}} = \frac{\sqrt{\sigma_{A_{LL}^{raw}}^2+r^2\sigma_{A^{BG}_{LL}}^2}}{1-r}. 
\label{eq:bsub}
\end{equation}

The spin asymmetry for each beam fill~\cite{foot1}
$A^{fill}_{LL}$ was calculated using Eq.~\ref{eq:a_ll}. 
For the $A^{fill}_{LL}$ error evaluation,
we considered only the $N_{++}$ and $N_{+-}$ statistical errors.
The resulting $A_{LL}$ was obtained 
after fitting a constant to all $A^{fill}_{LL}$'s. The fit 
$\chi^2_{fit}$ and a ``bunch shuffling'' technique were used to check 
the uncertainties assigned to $A_{LL}$. In each ``bunch shuffling'' we 
randomly assigned the helicity sign to every bunch crossing, keeping the 
balance between the number of bunches with correctly and inversely assigned 
helicities, so that the average polarization for each shuffled sample 
was nearly zero, and recalculated $A_{LL}$. 
The widths of the distributions of $A_{LL}$ values 
obtained in all bunch shuffles were consistent with 
errors assigned to $A_{LL}$ indicating that all non-correlated bunch-to-bunch and 
fill-to-fill systematic errors were much smaller than the
$\pi^{0}$ yield statistical errors. 

A number of systematic checks, including variation of photon identification criteria 
and mass window range for $\pi^0$s and background,  were performed to look 
for possible systematic effects on the measured $A_{LL}$ values. 
None were found. 

The double spin asymmetries between $(++)$ and $(--)$ 
and between $(+-)$ and $(-+)$ helicity configurations, as well as the
single spin asymmetries for each polarized beam 
($A_{L} = -\frac{\sigma_{+}-\sigma_{-}} {\sigma_{+}+\sigma_{-}}$) 
were evaluated. These measure parity violating asymmetries, if any. 
All of these asymmetries were consistent with zero.

\begin{table}
\caption{\label{tab:final} Double spin asymmetry for the raw signal 
($\pi^0+BG$), for the background ($BG$) and for $\pi^0$ background 
corrected; single spin asymmetry for $\pi^0$ background 
corrected; for the four $p_{T}$ bins, the mean $p_T$ of $\pi^0$'s being 
1.59, 2.39, 3.37 and 4.38~GeV/$c$, in four $p_T$ bins correspondingly . 
The numbers in brackets in the second column indicate the background 
contribution to the $\pi^{0}$ signal.}
\begin{ruledtabular} \begin{tabular}{ccccc}
$p_T$ & $A_{LL}^{raw}$ & $A_{LL}^{BG}$ & $A_{LL}^{\pi0}$ & $A_{L}^{\pi0}$  \\
(GeV/$c$) &  ($\%$)       &  ($\%$)       &  ($\%$)         &  ($\%$) \\
\hline
1-2    &  $-1.5 \pm 0.9$ (27\%) & $ 1.6 \pm 1.4$ & $-2.7 \pm 1.3$ & $-0.2 \pm 0.3$ \\
2-3    &  $-1.5 \pm 1.1$ (15\%) & $-3.0 \pm 2.4$ & $-1.3 \pm 1.3$ & $-0.1 \pm 0.3$ \\
3-4    &  $-1.8 \pm 2.5$ (9\%)  & $-2.4 \pm 6.8$ & $-1.7 \pm 2.8$ & $-0.3 \pm 0.6$ \\
4-5    &  $ 2.6 \pm 5.7$ (8\%)  & $  24 \pm 17 $ & $ 0.7 \pm 6.2$ & $-1.0 \pm 1.2$ \\
\end{tabular} \end{ruledtabular}
\end{table}

\begin{figure}
\includegraphics[width=0.9\linewidth]{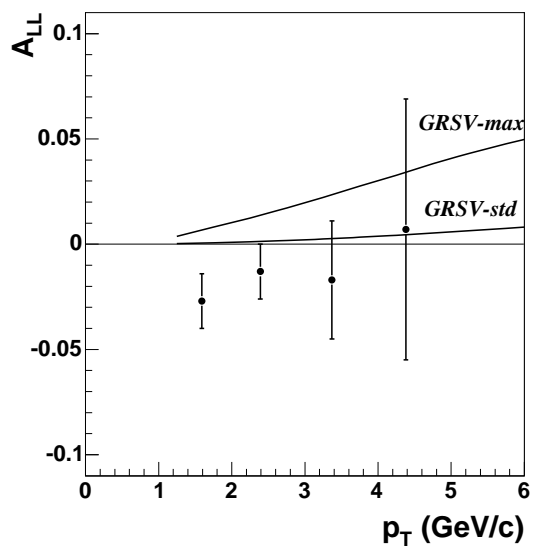}
\caption{\label{fig:all} The measured double spin asymmetry $A_{LL}^{\pi^{0}}$ 
versus mean $p_{T}$ of $\pi^0$'s in each bin. 
A scale uncertainty of $\pm65\%$ is not included.
Two theoretical calculations based on
NLO pQCD are also shown for comparison with the data (see text for details). 
}
\end{figure}

The results of the asymmetry are
presented in Table~\ref{tab:final} and Fig.~\ref{fig:all}.  
Systematic uncertainties for the
asymmetry measurement are negligible.  A total scale uncertainty 
of $\pm65\%$, from the correlated polarization analyzing power uncertainty, 
$\delta A_{N}^{pC}$, for
the two beams and the uncorrelated measurement uncertainties, is not shown.

Two theoretical curves based on NLO pQCD are shown in Fig.~\ref{fig:all},
representing different assumptions for the gluon polarization, with
GRSV-std using the best global fit to inclusive DIS data, and GRSV-max using a
gluon polarized distribution equal to the unpolarized distribution at
the input scale of $Q^2=0.6~{\rm GeV}^2$~\cite{theory, GRSV}.  
The range of theory curves reflects the uncertainty on gluon polarization 
from the inclusive DIS measurements~\cite{GRSV, BB, AAC}. 
The uncertainties from this measurement are similar in size to this 
range~\cite{theory3}. Both curves show positive $A_{LL}$.
It was recently argued that a negative $A_{LL}$
would be difficult to accommodate~\cite{theory2}.  
The present results are consistent
with the GRSV-std curve, with CL=16-20\%, for the range in polarization
scale uncertainty.  For GRSV-max, CL=0.02-5\%.  These confidence levels
do not include a theoretical uncertainty.  
In addition, the lowest $p_T$ data point, at $\langle p_T \rangle$=1.59
GeV/$c$, may have a significant soft-physics contribution.  The agreement
of the NLO pQCD calculation for the cross section~\cite{pi0_prl}
can only be checked
to within the uncertainties of the calculation, estimated to be a factor
two at this $p_T$~\cite{theory,theory2}.

In summary, the reported results of the double helicity asymmetries for
$\pi^0$ production begin to explore the gluon polarization in the proton,
using strongly interacting probes.
The observed asymmetry is small, and the
level of uncertainty of these data is comparable to that of
the theoretical calculations adopting different sets of gluon
polarizations deduced from global fits to the existing 
deep inelastic scattering data.



We thank the staff of the Collider-Accelerator Department, Magnet
Division, and Physics Department at BNL and the RHIC polarimetry group for
their vital contributions.  We thank W.~Vogelsang for informative
discussions.  We acknowledge support from the Department of Energy and NSF
(U.S.A.), MEXT and JSPS (Japan), CNPq and FAPESP (Brazil), NSFC (China),
IN2P3/CNRS, CEA, and ARMINES (France), BMBF, DAAD, and AvH (Germany), OTKA
(Hungary), DAE and DST (India), ISF (Israel), KRF and CHEP (Korea), RAS,
RMAE, and RMS (Russia), VR and KAW (Sweden), U.S. CRDF for the FSU,
US-Hungarian NSF-OTKA-MTA, and US-Israel BSF.


\clearpage


\begin{references}

\bibitem{DISsigma} EMC, J. Ashman {\em et al.}, Phys. Lett. {\bf B206} 
                   364 (1988), Nucl. Phys. {\bf B328}, 1 (1989);
              E. Hughes and R. Voss, Ann. Rev. Nucl. Part. Sci. {\bf 49}, 
              303 (1999). 
\bibitem{DISg1}SMC, B. Adeva {\em et al.}, Phys. Rev. {\bf D58}, 112002 (1998);
               E155, P. L. Anthony {\em et al.}, Phys. Lett. {\bf B 493}, 19 (2000).
\bibitem{DIShad} HERMES, A. Airapetian {\em et al.}, Phys. Rev. Lett. {\bf 84}, 2584 (2000); SMC, B. Adeva {\em et al.}, hep-ex/0402010.
\bibitem{E704} D.L.~Adams {\it et al.}, \PLB {\bf 261}, 197, (1991).
\bibitem{pi0_prl} S.S.~Adler {\it et al.}, \PRL {\bf 91}, 241803 (2003). 
\bibitem{bunce} G.~Bunce {\it et al.}, Ann. Rev. Nucl. Part. Sci. {\bf 50}, 525 (2000). 
\bibitem{pol} O.~Jinnouchi {\it et al.}, 
15th Int. Spin Physics Symposium
(SPIN 2002), AIP Conf.~Proc.~675: 817-825, 2003; 
Xth Workshop on High Energy Spin Physics (SPIN 2003), 
Dubna, Russia, Sep. 16-20, 2003.
\bibitem{tojo} J.~Tojo {\it et al.}, \PRL {\bf 89}, 052302 (2002).
\bibitem{trueman} T.L.~Trueman, hep-ph/0203013.
\bibitem{locpol} A.~Bazilevsky {\it et al.}, 
15th Int. Spin Physics Symposium 
(SPIN 2002), AIP Conf.~Proc.~675: 584-588, 2003.
\bibitem{nim_zdc} C.~Adler {\it et al.}, \NIMA {\bf 470}, 488 (2001).
\bibitem{sqrtform} G.G.~Ohlsen and P.W.~Keaton, Jr., \NIM {\bf 109}, 41 (1973).
\bibitem{nim_bbc} M.~Allen {\it et al.}, \NIMA {\bf 499}, 549 (2003).
\bibitem{nim_emc}L.~Aphecetche {\it et al.}, \NIMA {\bf 499}, 521 (2003).
\bibitem{nim_pc}K.~Adcox {\it et al.}, \NIMA {\bf 499}, 489, (2003).
\bibitem{foot1}Each fill is characterized by a constant polarization of 
the beams.
\bibitem{theory} B.~J\"ager {\it et al.}, \PRD {\bf 67}, 054005 (2003).
\bibitem{GRSV} M.~Gl\"uck {\it et al.}, \PRD {\bf 63}, 094005 (2001).
\bibitem{BB} J.~Bl\"{u}mlein and H.~B\"{o}ttcher, \NPB {\bf 636}, 225 (2002).
\bibitem{AAC} M.~Hirai {\em et al.}, Phys. Rev. {\bf D69}, 054021 (2004).
\bibitem{theory3} M.~Hirai and K.~Sudoh, hep-ph/0403102; 
RIKEN preprint RIKEN-AF-NP-456.
\bibitem{theory2} B.~J\"ager {\it et al.}, hep-ph/0310197, 
to be published in \PRL

\end{references}
\end{document}